# Assessment of AlGaN/AlN superlattices on GaN nanowires as active region of electron-pumped ultraviolet sources

I. Dimkou[1], A. Harikumar[2], F. Donatini[3], J. Lähnemann[4], M. I. den Hertog[3], C. Bougerol[3], E. Bellet-Amalric[2], N. Mollard[2], A. Ajay[2], G. Ledoux[5], S. T. Purcell[5], and E. Monroy[2]

[1] Univ. Grenoble-Alpes, CEA, LETI, F-38000 Grenoble

[2] Univ. Grenoble-Alpes, CEA-IRIG-PHELIQS, 17 av. des Martyrs, 38000 Grenoble, France.

[3] Univ. Grenoble-Alpes, CNRS-Institute Néel, 25 av. des Martyrs, 38000 Grenoble, France.

[4] Paul-Drude-Institut für Festkörperelektronik, Leibniz-Institut im Forschungsverbund Berlin e. V., Hausvogteiplatz 5-7, 10117 Berlin, Germany

[5] Univ. Lyon, Université Claude Bernard Lyon 1, CNRS, Institut Lumière Matière, 69622 Lyon, France

## Abstract

In this paper, we describe the design and characterization of 400-nm-long (88 periods) $Al_xGa_{1-x}N/AlN$ ($0 \leq x \leq 0.1$) quantum dot superlattices deposited on self-assembled GaN nanowires for application in electron-pumped ultraviolet sources. The optical performance of GaN/AlN superlattices on nanowires is compared with the emission of planar GaN/AlN superlattices with the same periodicity and thickness grown on bulk GaN substrates along the N-polar and metal-polar crystallographic axes. The nanowire samples are less sensitive to nonradiative recombination than planar layers, attaining internal quantum efficiencies (IQE) in excess of 60% at room temperature even under low injection conditions. The IQE remains stable for higher excitation power densities, up to 50 kW/cm². We demonstrate that the nanowire superlattice is long enough to collect the electron-hole pairs generated by an electron beam with an acceleration voltage $V_A$ = 5 kV. At such $V_A$, the light emitted from the nanowire ensemble does not show any sign of quenching under constant electron beam excitation (tested for an excitation power density around 8 kW/cm$^2$ over the scale of minutes). Varying the dot/barrier thickness ratio and the Al content in the dots, the nanowire peak emission can be tuned in the range from 340 to 258 nm.

## 1. Introduction

There is a particular demand for solid-state devices emitting in the ultraviolet (UV) range around 254 nm for application in disinfection and water purification [1–4]. Looking at the energy band gap and doping capabilities, AlGaN semiconducting alloys seem the natural material choice for this spectral range. However, in spite of intense research during the last decade [5,6], the efficiency of AlGaN-based light emitting diodes (LEDs) is relatively low (external quantum efficiency, EQE < 10% in the deep-UV) in comparison to visible LEDs or high-pressure mercury lamps. We remind that the EQE is the ratio of the number of photons emitted to the number of electrons injected in the device. The EQE is the product of the internal quantum efficiency (IQE), the carrier injection efficiency, and the light extraction efficiency. There are various reasons behind the low efficiency values. First, the scarcity of high-quality AlN substrates has motivated that most of the research is performed on non-native substrates, which in turn lead to high dislocation densities, which reduce the IQE in quantum wells. Second, the electrical injection is problematic due to the high activation energy of both donors and acceptors in AlGaN, and the asymmetric carrier mobility of electrons and holes. Attaining reasonable hole concentrations and ohmic contacts to the p-type region requires that the structure is terminated with p-GaN, which absorbs the deep-UV LED emission. As the Al content increases, AlGaN alloys present also problems for light extraction, since the polarization of the emitted light changes from surface-emitting (electric field component normal to the crystallographic $c$-axis for GaN) to edge-emitting (electric field parallel to the $c$-axis for AlN) [7].

An alternative technology for the fabrication of highly-efficient, eco-friendly UV lamps consists in mounting the active AlGaN chip in a vacuum tube, and to inject carriers using a cold cathode [8–10]. Such a device, which can be miniaturized to the millimetre size, would avoid the problems associated with electrical injection, since both electrons and holes are generated equally in the active region by impact ionization. Electron-pumped emission from GaN-based quantum wells has been reported [9,11–15], but the performance was mostly limited by the light extraction efficiency. Electron-pumped lasing around 350 nm has also been demonstrated [16,17].

The use of nanowire arrays as active media opens interesting perspectives for electron-pumped UV emitters. Semiconductor nanowires represent an innovative material approach with potential applications in the domains of electronics, optoelectronics, sensors and energy conversion [18–25]. The nanowire geometry offers the advantages of strain compliance, high crystalline quality, and the possibility of self-assembled growth on silicon substrates, which reduces the device production costs. From the optical point of view, the nanowire geometry spontaneously favours light extraction without additional processing requirements. In addition the relatively easy implementation of nano-objects displaying three-dimensional (3D) quantum confinement (quantum-dot-like) [26–29] results in high IQE at room temperature [30,31]. With a nanowire diameter around 50 nm the "quantum dot" nature of insertions in nanowires is often controversial, since the radial confinement is very week in comparison to the vertical confinement. Some authors argue that the objects look more like wells than like dots, and they are often referred to as "nanodisks". In this paper, we make the choice of referring to these objects as "quantum dots" based on their optical behaviour. In previous works we have reported studies of single GaN/AlN dots in a nanowire [26], and showed correlation measurements indicating antibunching [32], which is a signature of quantum dot behaviour. Furthermore, the strong reduction of non-radiative recombination, which results in photoluminescence lifetimes that remain constants in the 5-300 K temperature range [33], confirms that the presence of 3D confinement of carriers.

The synthesis of nanowire ensembles containing quantum dot superlattices faces some particular challenges. For application in electron-pumped devices it is important to attain a good control and reproducibility of the quantum dot geometry and composition in a superlattice that is hundreds-of-nanometres long (exceeding the penetration depth of the electrons). Unfortunately nanowires generally present large spectral dispersion [31,34] due to the variation of the dot geometry [35–37] and variable strains [35,36,38] along the growth axis, the appearance of interdiffusion at the heterointerfaces [31], and the structural perturbations introduced by nanowire coalescence [39].

In this contribution, we describe the design and characterization of N-polar $Al_xGa_{1-x}N/AlN$ ($0 \leq x \leq 0.1$) quantum dot superlattices deposited on self-assembled GaN nanowires for application in electron-pumped UV sources. The superlattice is long enough to collect the electron-hole pairs generated by an electron beam with an acceleration voltage $V_A = 5$ kV. Firstly, the optical performance of GaN/AlN quantum dot superlattices on nanowires is compared with the emission of planar GaN/AlN superlattices with the same periodicity and thickness grown on bulk GaN substrates along the N-polar and metal-polar crystallographic axes. Secondly, we demonstrate the adjustment of the nanowire peak emission in the range of 340 to 258 nm by varying the dot/barrier thickness ratio and the Al content in the dots.

## 2. Methods

Self-assembled N-polar nanowires were synthesized using plasma-assisted molecular beam epitaxy (PAMBE) on n-type Si(111) substrates on top of a low-temperature two-step AlN nucleation layer, as described elsewhere [40–42]. To improve the uniformity of the height across the substrate surface [43], the growth was started with a long ($\approx 900$ nm) GaN nanowire base grown under N-rich conditions (Ga/N flux ratio $\Phi_{Ga}/\Phi_N = 0.25$), at a substrate temperature $T_S = 810$ °C, and with a growth rate of $v_G = 330$ nm/h. This leads to a density of GaN nanowires around 6-8×10⁹ cm⁻², with a nanowire diameter of 30-50 nm. The growth continued with the deposition of the $Al_xGa_{1-x}N/AlN$ ($x = 0$, 0.05, 0.10) active region, which was 400 nm long (88 periods of quantum dots). The $Al_xGa_{1-x}N$ dots were grown using the same N-rich conditions as for the GaN base (Ga/N flux ratio = 0.25), and adding a flux of aluminium $\Phi_{Al} = x/v_G$, where $x$ is the targeted Al mole fraction. The Al mole fraction in the dots was intentionally kept low, $x \leq 0.1$, to prevent deformations of the nanowire morphology and reduce the effects of alloy inhomogeneity, observed for $x \geq 0.3$ [31]. The AlN sections were grown at stoichiometry ($\Phi_{Al}/\Phi_N = 1$). The complete heterostructure was synthesized at $T_S = 810$ °C without any growth interruption. A summary of the samples under study is presented in table 1. Some of the samples were repeated to assess the reproducibility of their growth.

In order to compare the performance of the nanowires with standard quantum wells, one of the samples (S1A) has been reproduced as a planar structure with similar thicknesses of well/dot, barrier and total active region. The sample was grown simultaneously on the Ga-face (P1G in table 1) and on the N-face (P1N in table 1) of free-standing n-type GaN substrates (resistivity at room temperature < 0.05 Ωcm, dislocation density < 5×10⁶ cm⁻²). Growth was performed at 720 °C, under Ga-rich conditions and without growth interruptions [44].

The structural quality was analysed by X-ray diffraction (XRD) in a Rigaku SmartLab diffractometer using a 2 bounce Ge(220) monochromator and a long plate collimator of 0.228° for the secondary optics. The as-grown nanowire ensemble was imaged using a Zeiss Ultra 55 field-emission scanning electron microscope (SEM). Detailed structural studies were conducted using high-resolution transmission electron microscopy (HR-TEM) and high-angle annular dark-field scanning transmission electron microscopy (HAADF-STEM) performed on a FEI Tecnai microscope and a probe-corrected FEI Titan Themis microscope, both operated at 200 kV.

Photoluminescence (PL) measurements under continuous-wave excitation were obtained by pumping with a frequency-doubled solid-state laser ($\lambda$ = 244 nm), with an optical power of $\approx 10\,\mu W$ focused on a spot with a diameter of $\approx 100\,\mu m$. PL measurements under pulsed excitation used a Nd-YAG laser (266 nm, 2 ns pulses, repetition rate of 8 kHz). In both cases, samples were mounted on a cold-finger cryostat, and the PL emission was collected by a Jobin Yvon HR460 monochromator equipped with a UV-enhanced charge-coupled device (CCD) camera.

Cathodoluminescence (CL) experiments were performed using a FEI Inspect F50 field-emission SEM equipped with a low-temperature Gatan stage to cool the sample down to 6 K, and with an IHR550 spectrometer. The beam spot diameter was $\approx 10\,nm$ on the focal point, the accelerating voltage was varied from 2 to 20 kV and the electron beam current was kept below 150 pA. Cross-sectional CL measurements were performed in a Zeiss Ultra-55 field-emission scanning electron microscope equipped with a Gatan MonoCL4 system and a cold stage, using an acceleration voltage of 5 kV and a current of about 1.3 nA. Spectral line-scans were recorded with an integration time of 0.5 s per spectrum on a cleaved cross-section and dispersed nanowires for samples P1G and S1A respectively. Finally, additional CL experiments were performed using a Kimball Physics EGPS-3212 electron gun operated in direct current mode, under normal incidence, with a beam spot diameter of 4±1 mm. The gun was operated with an acceleration voltage in the range of 3 to 10 kV, injecting up to 400 $\mu A$ of current. The CL emission arrived to an ANDOR ME-OPT-0007 UV-NIR light collector coupled with an ANDOR Shamrock500i spectrograph connected to an electron-multiplying CCD Newton 970 from ANDOR operated in conventional mode.

The electronic structure of samples S1A and S8 was modelled in 3D using the Nextnano³ 8-band **k·p** Schrödinger-Poisson equation solver [28], with the material parameters described in ref. [29]. For the $Al_xGa_{1-x}N$ alloys, all the bowing parameters were set to zero. The nanowires were defined as a regular hexagonal prism with minor radii being 27.5 nm. Along the axis, they consisted of a 150-nm-long GaN base section, followed by 15 periods of $Al_xGa_{1-x}N$/AlN quantum dots. Both the $Al_xGa_{1-x}N$ /AlN dots and the GaN base were laterally surrounded by a 5-nm-thick AlN shell, and the whole nanowire structure was enclosed in a rectangular prism of air, permitting elastic deformation. In a first step, the 3D strain distribution was calculated by minimizing the elastic energy assuming zero stress at the nanowire surface. This was followed by the calculation of the band diagram, taking both spontaneous and piezoelectric polarization into account. For such calculations, it was assumed that the Fermi level at the AlN/air sidewall interface is pinned 2.1 eV below the conduction band minimum [45]. Finally, the electronic levels and squared electron and hole wavefunctions were calculated in the quantum dot located in the middle of the $Al_xGa_{1-x}N$/AlN heterostructure.

## 3. Design

The aim of this study is to explore the possibility of using $Al_xGa_{1-x}N$/AlN quantum dot superlattices integrated along GaN nanowires as active media for electron-pumped UV emitters. The implementation of a superlattice of quantum dots in a wire seems a good strategy, since 3D carrier confinement grants a certain insensitivity to non-radiative processes [32], which results in high internal quantum efficiency at room temperature.

However, the semiconductor geometry and conductivity must be adapted to maximize the energy conversion under electron pumping.

A first specification is that the total length of the active region should be larger than the penetration depth of the impinging electron beam, $R_e$. A commonly used empirical expression for $R_e$ is [46]:

$$R_e = \frac{4.28 \times 10^{-6}}{\rho} V_A^{1.75} \qquad (1)$$

where $\rho$ is the material density in g/cm³ and $V_A$ is the acceleration voltage in kV. Applied to GaN ($\rho = 6.10\,g/cm^{-3}$) and AlN ($\rho = 3.26\,g/cm^{-3}$) assuming $V_A = 5$ kV, we obtain $R_e = 115$ nm and $R_e = 215$ nm, respectively. For a more precise estimation of $R_e$, Monte Carlo simulations of the electron beam interaction with GaN, Al$_{0.67}$Ga$_{0.33}$N and AlN were performed as a function of $V_A$, using the CASINO software [47], with the results depicted in figure 1. For $V_A = 5$ kV, the calculations predict values of $R_e$ around 175 nm, 224 nm, and 260 nm for GaN, Al$_{0.67}$Ga$_{0.33}$N and AlN, respectively [see figures 1(a), (b) and (c), respectively]. The penetration depth in Al$_{0.67}$Ga$_{0.33}$N is presented here because it reflects the average Al content in the active region of some of the heterostructures, as it will be explained later in this paper.

In view of these results, and accounting for a reduction of the effective material density in the nanowire ensemble with respect to bulk material, the length of our quantum dot superlattices was chosen to be 400 nm. To maximize the number of emitting centres, the nominal superlattice period was fixed at 4.5 nm, which means that the active region consists of 88 quantum dot periods. To adjust the spectral response, we varied the Al content in the dots and the dot/barrier thickness ratio, i.e. the total number of dots and the active region length remained constant.

To favour the charge evacuation during the electron pumping process the nanowires were grown on n-type Si(111) with a resistivity in the range of 0.001-0.005 Ωcm. The nanowire base and the dots were doped n-type with [Si] = 5×10$^{18}$ cm$^{-3}$ (value estimated from Hall effect measurements using the Van der Pauw method on planar Si-doped GaN layers).

## 4. Results and discussion

### 4.1 Nanowires vs. planar layers

In a first experiment, we have compared the structural and optical properties of GaN/AlN quantum dot superlattices on GaN nanowires (samples S1A and S1B in table 1) with planar structures with the same layer sequence grown along the [0001] (Ga-polar) or [000-1] (N-polar) crystallographic axis of GaN, starting from free-standing GaN substrates (samples P1G and P1N in table 1, respectively). An SEM image of S1A is displayed in figure 2(a), illustrating the high density of nanowires in the ensemble. Figures 2(b-c) present HAADF-STEM images of the GaN/AlN heterostructure (88 periods) in S1A, where the bright contrast is GaN and the darker contrast corresponds to AlN. Along the nanowire, the heterostructure exhibits a regular thickness of 1.5 ± 0.2 nm for GaN and 2.8 ± 0.2 nm of AlN, with the whole GaN/AlN quantum dot superlattice enveloped by an AlN shell. Along the superlattice the thickness of the GaN section increases by a maximum of 0.5 nm from the base to the end. It can be observed in zoomed images [figure 2(c)] that the thickness of the AlN shell decreases along the nanowire growth axis, being negligible at the top of the nanowires and having a maximum thickness of around 5 nm at the interface with the GaN base. The vertical cross section of the GaN dots is not rectangular but trapezoidal, with {1−102} facets towards the nanowire sidewalls.

Figure 2(d) depicts the integrated intensity of the HAADF-STEM image in (c) analysed along the turquoise and red rectangles outlined in the figure. From these profiles, the interface resulting from the deposition of AlN on GaN is systematically sharp at the scale of one atomic layer. On the contrary the GaN-on-AlN interface is widened by interdiffusion, extending 2-3 atomic layers. Note that the results are similar in the centre of the nanowire (data in turquoise) as closer to the sidewalls (data in red). If we focus on the initial stage of the growth of the heterostructure, illustrated in figure 2(e), there is a diffusion of Ga from the stem into the first two periods of the superlattice. Such Al-Ga intermixed regions have previously been reported at the beginning of the growth of GaN/AlN superlattices [31], and are explained by the strain-driven out-diffusion of Ga induced by the lattice mismatch between GaN and AlN.

As a comparison, figure 3(a) displays a HAADF-STEM image of a Ga-polar planar sample containing 88 periods of GaN/AlN (P1G), and figure 3(b) is a bright-field TEM off-axis (10° tilt from the [11−20] zone axis) image showing five periods in the centre of the superlattice. The structure presents sharp and symmetric hetero-interfaces at the atomic level, only disturbed by clearly identified atomic steps [marked by white arrows in figure 3(b)], characteristic of the step-flow layer-by-layer growth mode. The equivalent N-polar planar structure is shown in figures 3(c-d). The heterointerfaces are still relatively sharp and symmetric, but they extend to about two atomic layers.

The structural characteristics of the GaN/AlN superlattice were also investigated by high-resolution XRD. Figure 4(a) shows θ−2θ scans around the GaN (0002) reflection for samples S1A, P1G and P1N. In the nanowire sample, the angular location of the reflection from the GaN stem is shifted to higher angles with respect to relaxed GaN, which confirms that it is compressively strained by the AlN shell. From the inter-satellite distance, we extract the MQW periods listed in table 1, which are in good agreement with the nominal values. Using the Ga-polar sample as a reference, the N-polar structure presents broader reflections due to the larger inhomogeneity in thickness observed in the TEM images. In the case of nanowires, thickness fluctuations from wire to wire within the ensemble introduce an additional broadening factor. Nevertheless, the satellites of the MQW reflection are still well resolved, and their linewidth is comparable to the N-polar structure.

To get an idea of the impact of the structural properties on the optical characteristics we studied the PL emission of the three heterostructures S1A, P1G and P1N. Normalized spectra recorded at low temperature (5 K) are displayed in figure 4(b). The peak emission wavelength of the three samples is relatively close, in the 336-340 nm range. The spectra from planar samples present a multi-peak structure due to the monolayer thickness fluctuations in the quantum wells [48], with the Ga polar sample presenting narrower lines. In contrast the emission for the nanowire sample consists of a single spectral line, centred at 337 nm, with a large linewidth due to the geometry fluctuations in the nanowire ensemble.

The variation of the PL intensity as a function of temperature is presented in figure 4(c). The luminescence from the planar structures drops by more than two orders of magnitude when increasing the temperature from 5 K to 300 K. In contrast the emission intensity from the nanowires remains constant up to 100 K and remains at 63% of its maximum (low temperature) value at 300 K. This result confirms the relevance of carrier confinement in hindering non-radiative recombination.

The room-temperature internal quantum efficiency (IQE) of the nanostructures is often estimated as the ratio of the room-temperature and low-temperature integrated PL intensities:

$$IQE(300K) = \frac{I(300K)}{I(0K)} \qquad (2)$$

This estimation is based on the hypothesis that the PL intensity saturates at low temperature as a result of the carrier freeze out which prevents carriers from reaching defect-related non-radiative recombination centres. There is hence a risk of overestimation of the IQE if non-radiative recombination is active at low temperature, but also if the pumping intensity is high enough to saturate non-radiative recombination paths at high temperature.

To get an estimation of the IQE associated with the material properties, we tried to introduce as few extraneous perturbations as possible to straightforward linear response. Thus the measurements presented here were performed using a continuous-wave laser, and very low power density excitation (10 µW laser power, focused on a spot with a diameter of 100 µm). The results in figure 4(c) show that the luminescence from planar samples P1G and P1N clearly does not attain saturation at low temperature, so that the IQE calculated from these curves should be taken as an upper limit. On the contrary, all the nanowire samples in the study present a clear saturation of their PL for temperatures lower than 100 K, as shown for S1A in figure 4(c). Therefore, the ratio of their room-temperature and low-temperature PL intensities can be considered as a good estimation of their IQE. The obtained IQE values are summarized in table 1, together with the peak emission wavelength and linewidth at room temperature. In the case of S1A and S1B, the IQE at room temperature is 60-63%, orders of magnitude higher than the IQE of planar structures with the same layer sequence (P1G and P1N). The higher value of IQE in P1N in comparison to P1G is explained by the localization of carriers in the thickness fluctuations observed in the N-polar structure.

When the perturbation introduced by the excitation source is very small (low injection regime), the IQE values describe the material properties in terms of radiative and non-radiative processes. However, it is difficult to compare these data with the literature since most reported values are measured under pulsed excitation [49–58], using power densities in the range of 5-1000 kW/cm² [49–52,54,55], to emulate the carrier injection at LED operating conditions. In this pumping regime, the photo-generated carrier densities are higher than the doping level of the original structure (high injection) and nonradiative recombination paths are partially saturated. Therefore the obtained IQE is significantly higher than the low-injection value and depends on the excitation power density [49–52,55,59]. Maximum values of IQE are obtained for an excitation power density around 10-100 kW/cm² [52,55]. Note that in an electron-pumped UV lamp using an acceleration voltage of 5 kV and an injection current of 1 mA to irradiate a spot with a diameter of 1 mm, the excitation density would be below 1 kW/cm².

To explore the behaviour of the nanowire heterostructures under operating conditions, and compare with previous literature, we have measured PL as a function of the excitation power using a pulsed Nd-YAG laser. Measurements were performed at 6 K and at 300 K. Under high injection, the calculation of the IQE at room temperature must take into account the drop of the PL efficiency at low temperature due to the many-body effects induced by high-power excitation [55] so that

$$IQE(300K, P) = \frac{I(300K, P)}{I(0K, P)} \times \frac{I(0K, P)}{I(0K, low\,injection)} \qquad (3)$$

where $I(T, P)$ is the integrated PL intensity as a function of temperature and excitation power ($P$). The results for samples S1A, P1G and P1N are presented in figure 4(d). In the case of planar samples, the IQE increases with pulsed excitation. A maximum IQE around 3-4% is obtained for excitation in the range of 100-500 kW/cm². The values are lower than the IQE = 15-50% that can be found in the literature for this spectral range (measured by the same method) [52,55,59]. This difference can be explained by the choice of AlN/GaN as materials for barriers/wells, to be used as a reference for the nanowire heterostructures. In this case, the lattice mismatch is maximum in comparison with generally-used $Al_xGa_{1-x}N/Al_yGa_{1-y}N$ heterostructures. This leads to enhanced plastic relaxation of the misfit strain [60], which has an important effect on the maximum IQE [50]. Looking back to figure 4(d), in the case of quantum dots contained in nanowires, the IQE is stable under pulsed excitation up to around 50 kW/cm², and then decreases slightly.

We have compared the optical characterization results with theoretical calculations of the electronic structure. A schematic description of the simulated structure is presented in figure 5(a), where the GaN dots are modelled as regular hexagonal prisms. To assess the effect of the sidewall facets observed in figure 2(c), the structure was also modelled considering a reduction of the prism radius along the growth axis, as illustrated in figure 5(b). The band profile of 3 quantum dots in the centre of the superlattice, taken at the radial centre of the nanowire [dashed lines in figures 5(a) and (b)], is displayed in figures 5(c) and (d), for calculations with and without sidewall facets, respectively. The squared wavefunctions of the ground electron and hole levels depicted in the figure represent in-plane integrated values. The effect of the polarization in the bands, rendering the characteristic saw-tooth profile of GaN, is clearly visible. The internal electric field in the dots shifts the electron wavefunction towards the nanowire stem and the hole wavefunction towards the cap. However, the separation of electron and hole along the growth axis remains relatively small due to the small quantum dot height. The electron-hole transition at room temperature is predicted to occur at 340 nm and 349 nm, as a result of the calculations in figures 5(a,c) and 5(b,d), respectively. These results are very close to the experimental value of 340 nm (room temperature, 31 nm linewidth) in sample S1A (i.e. the sample analysed by HAADF-STEM in figure 2).

If we compare the results of the calculations for the structures in figures 5(a) and (b), the presence of the facets in the dots does not have a major effect on the strain in the centre of the nanowire (see values in table 2). However, it has a drastic effect on the in-plane distribution of the electron and hole wavefunctions, depicted on the right side of figures 5(c) and (d). In the structure without facets [figure 5(c)] the maximum of the electron wavefunction is located in the centre of the nanowire, whereas the hole is shifted towards the sidewalls. In the structure with facets the radial distribution of carriers is opposite, with the hole wavefunction in the centre of the nanowire and the electron close to the sidewalls. The radial separation of the electron and hole wavefunctions is known to exist in GaN nanowires and its effects on the carrier lifetime have been measured by time-resolved PL [33]. In homogeneous nanowires (without heterostructure) the separation occurs due to the Fermi level pinning at the nanowire sidewalls [61]. However in heterostructured nanowires the separation is enhanced due to a radial electric field that appears because of the radial gradient of strain and the sheer component of the strain associated to elastic relaxation at the surface [33]. Figures 5(e) and (f) represent the in-plane electric field along the [11−20] axis, measured at the point along the growth axis where the electron and hole wavefunctions reach their maxima. In the structure without facets [figure 5(e)] the field points towards the sidewalls, with a maximum value of 37 kV/cm. Even if this value is much smaller than the electric field along the growth axis (7.3 MV/cm in the dots), it is enough to force a radial separation of electron and hole. However it should be noted that the calculations do not take into account the Coulomb attraction, so that the radial separation should be smaller than the calculated result. The presence of the facets in the quantum dots changes the strain distribution, which leads to an inversion of the radial electric field, now more intense (maximum of 250 kV/cm) and pointing towards the centre of the wire [see figure 5(f)], which results in a different radial distribution of charges. In spite of these radial changes the emission wavelength does not vary significantly since it is dominated by the axial electric field, which is one order of magnitude more intense.

To better understand the origin of the spectral broadening of the nanowire samples, and the presence of multiple peaks in the planar layers, we investigated P1G and S1A using CL spectral line-scans on the cross-section of the samples, recorded at 7 K. The results are depicted in figures 6(a) and (b). In the case of the planar structure [figure 6(a)], analysing from bottom to top, the narrow emission from the GaN substrate is observed around 356 nm. The line remains visible when the excitation shifts into the superlattice (position = 0-0.4 μm in the graph), which points to a good vertical diffusion of carriers, facilitated by the orientation of the polarization-induced internal electric field for this Ga-polar sample. The emission from the superlattice presents two distinct peaks, which is consistent with the PL measurements in figure 4(b). The spectral separation of the two peaks corresponds to a change in quantum well thickness of 1 monolayer, which is in line with the atomic steps at the top interface of the quantum wells identified in HAADF-STEM [see figure 3(b)]. Towards the surface these peaks are slightly shifted to shorter wavelengths as outlined by the white dashed lines in the figure. The small value of the total shift, around 1 nm along the superlattice thickness, points to strain relaxation.

If we turn to the behaviour of the nanowire structure [figure 6(b)], the emission from the GaN stem around 350 nm is visible in the 0.4-0.5 μm position range. The fact that the emission is no longer observed when exciting the superlattice (position = 0-0.4 μm) is explained by the N-polarity of the nanowires. In this crystallographic orientation the internal electric field pushes the electrons away from the interface between the stem and the superlattice. The emission from the superlattice shifts from 320 to 344 nm, following the dashed white line in the figure, when the excitation moves from the stem to the top of the nanowire. Note that in individual spectra at a specific position, the emission linewidth is comparable to that of the planar superlattice. Therefore the spectral broadening of the luminescence in figure 4(b) is mostly due to the shift of the emission along the growth axis. Looking at theoretical calculations for GaN/AlN superlattices in nanowires [62] this shift can be explained by a difference in the quantum dot height of 1-2 atomic layers ($\approx$ 0.3-0.5 nm) along the growth axis, consistent with STEM observations.

To study the penetration depth of the electron beam under operating conditions, samples P1G and S1A were characterized by CL (excitation normal to the surface in the SEM) as a function of the acceleration voltage, with the results described in figure 7. Figure 7(a) shows that the superlattice presents a multi-peak emission as described by the PL studies [figure 4(b)]. Luminescence from the GaN substrate (at 357 nm) is only visible for acceleration voltages higher than 10 kV. This result is consistent with the calculations in figure 1(b), which show that acceleration voltages higher than 7.5 kV are required for electrons to generate electron hole pairs in the substrate. Note that the $Al_xGa_{1-x}N$ layer modelled in figure 1(b) presents the same average aluminium mole fraction as the active region of samples P1G and S1A. In the case of the nanowire sample [figure 7(b)] the emission spectra are dominated by a line around 340 nm, assigned to carrier recombination in the quantum dot superlattice. The emission from the stem becomes visible for acceleration voltages above 5.5 kV, which points to a higher penetration depth in comparison with planar layers, which is justified by the reduced average material density in the nanowire ensemble.

CL studies of GaN nanowires have been previously reported [34,63–65] with some groups observing a reversible degradation of the emission during the experiments [63,65] which they attributed to charge accumulation and trapping at the nanowire surface. The CL quenching was reduced in the case of nanowires with an AlN shell [63]. The stability of our samples under electron irradiation was studied by measuring the evolution of the CL intensity with time. The experiment was performed at room temperature with an acceleration voltage $V_A = 5$ kV, the injected current was 25 pA and the beam was concentrated in a 20 nm spot (i.e. excitation power density around 8 kW/cm$^2$ at the impact point). Figure 7(c) shows the evolution of the emission from P1G and S1A over a period of 300 seconds. In contrast to refs. [63,65], we did not observe any degradation of the CL emission. The stability of the samples is probably due to the relatively thick AlN shell that embeds the quantum dots and renders them insensitive to recombination at the nanowire surface.

Finally, to assess the influence of the electron beam current on the CL emission, the nanowire sample (S1A) was pumped with an electron gun operated in direct current mode, under normal incidence, with a beam spot diameter of 4±1 mm, and keeping $V_A = 5$ kV. As shown in figure 7(b), the integrated emission of the sample scales linearly with the injected current, without any indication of saturation up to 400 µA.

### 4.2 Towards shorter wavelengths

In order to shift the peak emission wavelength towards the spectral region efficient for disinfection (around 260-270 nm), we have first explored the effect of reducing the size of the GaN dots in the heterostructure from 1.5 nm to 0.65 nm, while keeping the same number of periods and total thickness of the heterostructure. Note that 0.65 nm is approximately 2.5 monolayers. The normalized room-temperature PL emission from these samples (S1B, S4, S5 and S7A) is displayed in figure 8 (green spectra), and the peak emission wavelength is summarized in table 1. The emission shifts from 340-336 nm (samples S1A-S1B) to 265-277 nm (samples S7A-S7B). However, the blue shift comes with a monotonous decrease of the IQE at room temperature measured in the low-injection regime, which evolves from 63% to ≈ 30%, as illustrated in the inset of figure 8.

To shift the emission further into the deep ultraviolet we have incorporated aluminium in the dots. The result is depicted in figure 8 (pink spectra). For 10% of Al in the dots, the room temperature emission is still dominated by a single spectral line that can be tuned from 331 nm (S2, 1.5 nm well thickness) to 258 nm (S8, 0.65 nm well thickness). Like in the case of GaN dots a decrease of the IQE and an increase of the relative linewidth is observed as the emission shifts to shorter wavelengths (see table 1). However, if we compare samples with GaN and $Al_{0.1}Ga_{0.9}N$ dots emitting at the same wavelength, the IQE is roughly the same (see inset of figure 8). Therefore, we conclude that the presence of Al in the dots does not introduce any additional degradation of the optical properties.

We have performed an analysis of the structural properties at the nanometre scale for sample S8, i.e. the structure emitting at the shortest wavelength, with the results illustrated in figure 9. An SEM image of S8 displayed in figure 9(a), does not show significant differences in shape or density with respect to the GaN/AlN samples [see figure 2(a)]. Figures 9(b-d) present HAADF-STEM images of the complete heterostructure and zoomed images of the last and first AlGaN/AlN periods, respectively. Bright contrast corresponds to Ga-rich areas and darker contrast represents Al-rich areas. If we focus on the initial stage of the growth of the heterostructure, illustrated in figure 9(d), the strain-driven diffusion of Ga from the stem into the superlattice is also visible in this case and it extends around 1.5 periods. The thickness of the dots and barriers is homogeneous along the nanowire (0.7-1 nm dots and 3.5-3.7 nm barriers). The image contrast points to a slight increase of the Ga content in the dots along the growth axis but this might be partially due to the increasing thickness of the AlN shell when moving from the top of the wire to the GaN stem, or to a slight bending of the nanowire crystal resulting in the heterostructure being better oriented with respect to the electron beam in the top part of the wire.

## 5. Conclusion

In summary, we present a feasibility study of nanowire arrays as active media for electron-pumped UV emitters. The nanostructures under study consist of 400-nm-long $Al_xGa_{1-x}N/AlN$ ($0 \leq x \leq 0.1$) quantum dot superlattices grown on self-assembled GaN nanowires by PAMBE on n-type Si(111) wafers. By growing the $Al_xGa_{1-x}N$ sections under N-rich conditions and the AlN sections under stoichiometric conditions, we obtain quantum dots with highly homogeneous dimensions along the whole superlattice. When compared with the emission of planar samples with the same periodicity and thickness, GaN/AlN nanowire structures are less sensitive to nonradiative recombination, attaining IQE higher than 60% at room temperature, even under low injection conditions. Using pulsed optical excitation we demonstrate that the IQE remains stable for excitation power densities up to 50 kW/cm². Experiments under electron pumping show that the nanowire superlattice is long enough to collect the electron-hole pairs generated by an electron beam with an acceleration voltage $V_A$ = 5 kV. By varying the dot/barrier thickness ratio and the Al content in the dots, we demonstrate the adjustment of the nanowire peak emission in the range from 340 to 258 nm.

## Acknowledgements

This work is supported by the French National Research Agency (ANR) via the UVLASE program (ANR-18-CE24-0014) and the GaNEX program (ANR-11-LABX-0014), and by the Auvergne-Rhône-Alpes region (grant PEAPLE). This project has also received funding from the European Research Council under the European Union's H2020 Research and Innovation programme via the e-See project (grant #758385). We also acknowledge technical support from F. Jourdan, Y. Curé and Y. Genuist. We benefited from the access to the technological platform NanoCarac of CEA-Minatech Grenoble in collaboration with the IRIG-LEMMA group.

# References

[1]    Würtele M A, Kolbe T, Lipsz M, Külberg A, Weyers M, Kneissl M and Jekel M 2011 Application of GaN-based ultraviolet-C light emitting diodes – UV LEDs – for water disinfection *Water Res.* **45** 1481–9

[2]    Beck S E, Ryu H, Boczek L A, Cashdollar J L, Jeanis K M, Rosenblum J S, Lawal O R and Linden K G 2017 Evaluating UV-C LED disinfection performance and investigating potential dual-wavelength synergy *Water Res.* **109** 207–16

[3]    Chen J, Loeb S and Kim J-H 2017 LED revolution: fundamentals and prospects for UV disinfection applications *Environ. Sci. Water Res. Technol.* **3** 188–202

[4]    Kneissl M, Seong T-Y, Han J and Amano H 2019 The emergence and prospects of deep-ultraviolet light-emitting diode technologies *Nat. Photonics* **13** 233–44

[5]    Kneissl M and Rass J 2016 A Brief Review of III-Nitride UV Emitter Technologies and Their Applications *III-Nitride Ultraviolet Emitters* vol 227 (Cham: Springer International Publishing) pp 1–25

[6]    Nagasawa Y and Hirano A 2018 A Review of AlGaN-Based Deep-Ultraviolet Light-Emitting Diodes on Sapphire *Appl. Sci.* **8** 1264

[7]    Northrup J E, Chua C L, Yang Z, Wunderer T, Kneissl M, Johnson N M and Kolbe T 2012 Effect of strain and barrier composition on the polarization of light emission from AlGaN/AlN quantum wells *Appl. Phys. Lett.* **100** 021101

[8]    Saito Y, Hamaguchi K, Mizushima R, Uemura S, Nagasako T, Yotani J and Shimojo T 1999 Field emission from carbon nanotubes and its application to cathode ray tube lighting elements *Appl. Surf. Sci.* **146** 305–11

[9]    Oto T, Banal R G, Kataoka K, Funato M and Kawakami Y 2010 100 mW deep-ultraviolet emission from aluminium-nitride-based quantum wells pumped by an electron beam *Nat. Photonics* **4** 767–70

[10]   Herbert J. Reich 2013 *Principles of Electron Tubes* (Literary Licensing, LLC)

[11]   Matsumoto T, Iwayama S, Saito T, Kawakami Y, Kubo F and Amano H 2012 Handheld deep ultraviolet emission device based on aluminum nitride quantum wells and graphene nanoneedle field emitters *Opt. Express* **20** 24320

[12]   Shimahara Y, Miyake H, Hiramatsu K, Fukuyo F, Okada T, Takaoka H and Yoshida H 2011 Fabrication of Deep-Ultraviolet-Light-Source Tube Using Si-Doped AlGaN *Appl. Phys. Express* **4** 042103

[13]   Ivanov S V, Jmerik V N, Nechaev D V, Kozlovsky V I and Tiberi M D 2015 E-beam pumped mid-UV sources based on MBE-grown AlGaN MQW: Mid-UV sources based on MBE-grown AlGaN MQW *Phys. Status Solidi A* **212** 1011–6

[14]   Tabataba-Vakili F, Wunderer T, Kneissl M, Yang Z, Teepe M, Batres M, Feneberg M, Vancil B and Johnson N M 2016 Dominance of radiative recombination from electron-beam-pumped deep-UV AlGaN multi-quantum-well heterostructures *Appl. Phys. Lett.* **109** 181105

[15]   Wang Y, Rong X, Ivanov S, Jmerik V, Chen Z, Wang H, Wang T, Wang P, Jin P, Chen Y, Kozlovsky V, Sviridov D, Zverev M, Zhdanova E, Gamov N, Studenov V, Miyake H, Li H, Guo S, Yang X, Xu F, Yu T, Qin Z, Ge W, Shen B and Wang X 2019 Deep Ultraviolet Light Source from Ultrathin GaN/AlN MQW Structures with Output Power Over 2 Watt *Adv. Opt. Mater.* **7** 1801763

[16]   Hayashi T, Kawase Y, Nagata N, Senga T, Iwayama S, Iwaya M, Takeuchi T, Kamiyama S, Akasaki I and Matsumoto T 2017 Demonstration of electron beam laser excitation in the UV range using a GaN/AlGaN multiquantum well active layer *Sci. Rep.* **7** 2944

[17]   Wunderer T, Jeschke J, Yang Z, Teepe M, Batres M, Vancil B and Johnson N 2017 Resonator-Length Dependence of Electron-Beam-Pumped UV-A GaN-Based Lasers *IEEE Photonics Technol. Lett.* **29** 1344–7

[18]   Gudiksen M S, Lauhon L J, Wang J, Smith D C and Lieber C M 2002 Growth of nanowire superlattice structures for nanoscale photonics and electronics *Nature* **415** 617–20

[19]   Lu W, Xie P and Lieber C M 2008 Nanowire Transistor Performance Limits and Applications *IEEE Trans. Electron Devices* **55** 2859–76

[20]   Krogstrup P, Jørgensen H I, Heiss M, Demichel O, Holm J V, Aagesen M, Nygard J and Fontcuberta i Morral A 2013 Single-nanowire solar cells beyond the Shockley–Queisser limit *Nat. Photonics* **7** 306–10

[21]   Xu Y, Gong T and Munday J N 2015 The generalized Shockley-Queisser limit for nanostructured solar cells *Sci. Rep.* **5** 13536

[22]   Hochbaum A I and Yang P 2010 Semiconductor Nanowires for Energy Conversion *Chem. Rev.* **110** 527–46

[23]   Cui Y 2001 Nanowire Nanosensors for Highly Sensitive and Selective Detection of Biological and Chemical Species *Science* **293** 1289–92

[24]   Spies M and Monroy E 2019 Nanowire photodetectors based on wurtzite semiconductor heterostructures *Semicond. Sci. Technol.* **34** 053002

[25]   Barrigón E, Heurlin M, Bi Z, Monemar B and Samuelson L 2019 Synthesis and Applications of III–V Nanowires *Chem. Rev.* **119** 9170–220

[26]   Renard J, Songmuang R, Bougerol C, Daudin B and Gayral B 2008 Exciton and Biexciton Luminescence from Single GaN/AlN Quantum Dots in Nanowires *Nano Lett.* **8** 2092–6

[27]   Nguyen H P T, Cui K, Zhang S, Fathololoumi S and Mi Z 2011 Full-color InGaN/GaN dot-in-a-wire light emitting diodes on silicon *Nanotechnology* **22** 445202

[28]   Holmes M J, Choi K, Kako S, Arita M and Arakawa Y 2014 Room-Temperature Triggered Single Photon Emission from a III-Nitride Site-Controlled Nanowire Quantum Dot *Nano Lett.* **14** 982–6

[29]   Gačević Ž, Vukmirović N, García-Lepetit N, Torres-Pardo A, Müller M, Metzner S, Albert S, Bengoechea-Encabo A, Bertram F, Veit P, Christen J, González-Calbet J M and Calleja E 2016 Influence of composition, strain, and electric field anisotropy on different emission colors and recombination dynamics from InGaN nanodisks in pencil-like GaN nanowires *Phys. Rev. B* **93** 125436

[30]   Jahangir S, Mandl M, Strassburg M and Bhattacharya P 2013 Molecular beam epitaxial growth and optical properties of red-emitting (λ = 650 nm) InGaN/GaN disks-in-nanowires on silicon *Appl. Phys. Lett.* **102** 071101

[31]   Himwas C, Hertog M den, Dang L S, Monroy E and Songmuang R 2014 Alloy inhomogeneity and carrier localization in AlGaN sections and AlGaN/AlN nanodisks in nanowires with 240–350 nm emission *Appl. Phys. Lett.* **105** 241908

[32]   Renard J, Kandaswamy P K, Monroy E and Gayral B 2009 Suppression of nonradiative processes in long-lived polar GaN/AlN quantum dots *Appl. Phys. Lett.* **95** 131903

[33]   Beeler M, Lim C B, Hille P, Bleuse J, Schörmann J, de la Mata M, Arbiol J, Eickhoff M and Monroy E 2015 Long-lived excitons in GaN/AlN nanowire heterostructures *Phys. Rev. B* **91** 205440

[34]   Zagonel L F, Mazzucco S, Tencé M, March K, Bernard R, Laslier B, Jacopin G, Tchernycheva M, Rigutti L, Julien F H, Songmuang R and Kociak M 2011 Nanometer Scale Spectral Imaging of Quantum Emitters in Nanowires and Its Correlation to Their Atomically Resolved Structure *Nano Lett.* **11** 568–73

[35]   Rigutti L, Teubert J, Jacopin G, Fortuna F, Tchernycheva M, De Luna Bugallo A, Julien F H, Furtmayr F, Stutzmann M and Eickhoff M 2010 Origin of energy dispersion in Al(x)Ga(1−x)N/GaN nanowire quantum disks with low Al content *Phys. Rev. B* **82** 235308

[36]   Furtmayr F, Teubert J, Becker P, Conesa-Boj S, Morante J R, Chernikov A, Schäfer S, Chatterjee S, Arbiol J and Eickhoff M 2011 Carrier confinement in GaN/AlxGa1−xN nanowire heterostructures (0<x≤1) *Phys. Rev. B* **84** 205303

[37]   Carnevale S D, Yang J, Phillips P J, Mills M J and Myers R C 2011 Three-Dimensional GaN/AlN Nanowire Heterostructures by Separating Nucleation and Growth Processes *Nano Lett.* **11** 866–71

[38]   Rivera C, Jahn U, Flissikowski T, Pau J, Muñoz E and Grahn H T 2007 Strain-confinement mechanism in mesoscopic quantum disks based on piezoelectric materials *Phys. Rev. B* **75** 045316

[39]   Zagonel L F, Rigutti L, Tchernycheva M, Jacopin G, Songmuang R and Kociak M 2012 Visualizing highly localized luminescence in GaN/AlN heterostructures in nanowires *Nanotechnology* **23** 455205

[40]   Ajay A, Lim C B, Browne D A, Polaczynski J, Bellet-Amalric E, den Hertog M I and Monroy E 2017 Intersubband absorption in Si- and Ge-doped GaN/AlN heterostructures in self-assembled nanowire and 2D layers *Phys. Status Solidi B* **254** 1600734

[41]   Musolino M, Tahraoui A, Fernández-Garrido S, Brandt O, Trampert A, Geelhaar L and Riechert H 2015 Compatibility of the selective area growth of GaN nanowires on AlN-buffered Si substrates with the operation of light emitting diodes *Nanotechnology* **26** 085605

[42]   Dimkou I, Harikumar A, Ajay A, Donatini F, Bellet-Amalric E, Grenier A, den Hertog M I, Purcell S T and Monroy E 2019 Design of AlGaN/AlN Dot-in-a-wire Heterostructures for Electron-Pumped UV Emitters *Phys. Status Solidi A* pssa.201900714

[43]   Sabelfeld K K, Kaganer V M, Limbach F, Dogan P, Brandt O, Geelhaar L and Riechert H 2013 Height self-equilibration during the growth of dense nanowire ensembles: Order emerging from disorder *Appl. Phys. Lett.* **103** 133105

[44]   Kandaswamy P K, Guillot F, Bellet-Amalric E, Monroy E, Nevou L, Tchernycheva M, Michon A, Julien F H, Baumann E, Giorgetta F R, Hofstetter D, Remmele T, Albrecht M, Birner S and Dang L S 2008 GaN∕AlN short-period superlattices for intersubband optoelectronics: A systematic study of their epitaxial growth, design, and performance *J. Appl. Phys.* **104** 093501

[45]   Reddy P, Bryan I, Bryan Z, Guo W, Hussey L, Collazo R and Sitar Z 2014 The effect of polarity and surface states on the Fermi level at III-nitride surfaces *J. Appl. Phys.* **116** 123701

[46]   Leamy H J 1982 Charge collection scanning electron microscopy *J. Appl. Phys.* **53** R51–80

[47]   Hovington P, Drouin D and Gauvin R 2006 CASINO: A new monte carlo code in C language for electron beam interaction -part I: Description of the program *Scanning* **19** 1–14

[48]   Tchernycheva M, Nevou L, Doyennette L, Julien F H, Guillot F, Monroy E, Remmele T and Albrecht M 2006 Electron confinement in strongly coupled GaN∕AlN quantum wells *Appl. Phys. Lett.* **88** 153113

[49]   Shatalov M, Yang J, Sun W, Kennedy R, Gaska R, Liu K, Shur M and Tamulaitis G 2009 Efficiency of light emission in high aluminum content AlGaN quantum wells *J. Appl. Phys.* **105** 073103

[50]   Ban K, Yamamoto J, Takeda K, Ide K, Iwaya M, Takeuchi T, Kamiyama S, Akasaki I and Amano H 2011 Internal Quantum Efficiency of Whole-Composition-Range AlGaN Multiquantum Wells *Appl. Phys. Express* **4** 052101

[51]   Bryan Z, Bryan I, Xie J, Mita S, Sitar Z and Collazo R 2015 High internal quantum efficiency in AlGaN multiple quantum wells grown on bulk AlN substrates *Appl. Phys. Lett.* **106** 142107

[52]   Murotani H, Akase D, Anai K, Yamada Y, Miyake H and Hiramatsu K 2012 Dependence of internal quantum efficiency on doping region and Si concentration in Al-rich AlGaN quantum wells *Appl. Phys. Lett.* **101** 042110

[53]   Liao Y, Thomidis C, Kao C and Moustakas T D 2011 AlGaN based deep ultraviolet light emitting diodes with high internal quantum efficiency grown by molecular beam epitaxy *Appl. Phys. Lett.* **98** 081110

[54]   Banal R G, Funato M and Kawakami Y 2011 Extremely high internal quantum efficiencies from AlGaN/AlN quantum wells emitting in the deep ultraviolet spectral region *Appl. Phys. Lett.* **99** 011902

[55]   Mickevičius J, Tamulaitis G, Shur M, Shatalov M, Yang J and Gaska R 2012 Internal quantum efficiency in AlGaN with strong carrier localization *Appl. Phys. Lett.* **101** 211902

[56]   Bhattacharyya A, Moustakas T D, Zhou L, Smith David J and Hug W 2009 Deep ultraviolet emitting AlGaN quantum wells with high internal quantum efficiency *Appl. Phys. Lett.* **94** 181907

[57]   Hao G-D, Tamari N, Obata T, Kinoshita T and Inoue S 2017 Electrical determination of current injection and internal quantum efficiencies in AlGaN-based deep-ultraviolet light-emitting diodes *Opt. Express* **25** A639

[58]   Dong P, Yan J, Zhang Y, Wang J, Zeng J, Geng C, Cong P, Sun L, Wei T, Zhao L, Yan Q, He C, Qin Z and Li J 2014 AlGaN-based deep ultraviolet light-emitting diodes grown on nano-patterned sapphire substrates with significant improvement in internal quantum efficiency *J. Cryst. Growth* **395** 9–13

[59]   Frankerl C, Hoffmann M P, Nippert F, Wang H, Brandl C, Tillner N, Lugauer H-J, Zeisel R, Hoffmann A and Davies M J 2019 Challenges for reliable internal quantum efficiency determination in AlGaN-based multi-quantum-well structures posed by carrier transport effects and morphology issues *J. Appl. Phys.* **126** 075703

[60]   Kandaswamy P K, Bougerol C, Jalabert D, Ruterana P and Monroy E 2009 Strain relaxation in short-period polar GaN/AlN superlattices *J. Appl. Phys.* **106** 013526

[61]   Calarco R, Marso M, Richter T, Aykanat A I, Meijers R, v.d. Hart A, Stoica T and Lüth H 2005 Size-dependent Photoconductivity in MBE-Grown GaN−Nanowires *Nano Lett.* **5** 981–4

[62]   Ajay A, Lim C B, Browne D A, Polaczyński J, Bellet-Amalric E, Bleuse J, den Hertog M I and Monroy E 2017 Effect of doping on the intersubband absorption in Si- and Ge-doped GaN/AlN heterostructures *Nanotechnology* **28** 405204

[63]   Lähnemann J, Flissikowski T, Wölz M, Geelhaar L, Grahn H T, Brandt O and Jahn U 2016 Quenching of the luminescence intensity of GaN nanowires under electron beam exposure: impact of C adsorption on the exciton lifetime *Nanotechnology* **27** 455706

[64]   Lim S K, Brewster M, Qian F, Li Y, Lieber C M and Gradečak S 2009 Direct Correlation between Structural and Optical Properties of III−V Nitride Nanowire Heterostructures with Nanoscale Resolution *Nano Lett.* **9** 3940–4

[65]   Robins L H, Bertness K A, Barker J M, Sanford N A and Schlager J B 2007 Optical and structural study of GaN nanowires grown by catalyst-free molecular beam epitaxy. II. Sub-band-gap luminescence and electron irradiation effects *J. Appl. Phys.* **101** 113506

**Table 1.** Structural and optical characteristics of the samples under study: Nominal thickness of the AlN barriers ($t_B$) and of the Al$_x$Ga$_{1-x}$N wells ($t_W$), Al concentration in the wells ($x$), MQW period measured by XRD, peak emission wavelength at room temperature (in the case of multiple peaks, the dominant peak appears in bold fonts), internal quantum efficiency (IQE) at room temperature (measurement under low-injection conditions), and full width at half maximum (FWHM) of the emission at room temperature.

| Sample | $t_W$ (nm) | $t_B$ (nm) | $x$ | XRD Period (nm) | Peak emission (nm) | IQE (%) | FWHM (nm) |
|---|---|---|---|---|---|---|---|
| S1A | 1.5 | 3.0 | 0 | 4.4±0.1 | 340 | 63 | 31 |
| S1B | | | | | 336 | 60 | 30 |
| S2 | 1.5 | 3.0 | 0.05 | 4.4±0.1 | 331 | 34 | 27 |
| S3 | 1.5 | 3.0 | 0.1 | 4.6±0.1 | 335 | 42 | 29 |
| S4 | 1.0 | 4.0 | 0 | 5.0±0.1 | 324 | 55 | 25 |
| S5 | 0.75 | 3.75 | 0 | 4.4±0.1 | 296 | 48 | 22 |
| S6 | 0.75 | 3.75 | 0.1 | 4.3±0.1 | 286 | 44 | 16 |
| S7A | 0.65 | 3.85 | 0 | | 277 | 29 | 23 |
| S7B | | | | 4.1±0.1 | 265 | 31 | 36 |
| S8 | 0.65 | 3.85 | 0.1 | 4.1±0.1 | 258 | 22 | 34 |
| P1G | 1.5 | 3.0 | 0 | 4.4±0.1 | 329, **343** | <0.05 | 9, **11** |
| P1N | 1.5 | 3.0 | 0 | 4.1±0.1 | 329, 342, **367** | <0.4 | 9, 13, **24** |

**Table 2.** Some results of the theoretical calculations described in figure 5: in-plane and out-of-plane strain at the centre of GaN dots ($\varepsilon_a^{QD}$ and $\varepsilon_c^{QD}$, respectively) and at the centre of the barrier ($\varepsilon_a^B$ and $\varepsilon_c^B$, respectively), and emission wavelength at room temperature ($\lambda_{RT}$).

| Simulation | $\varepsilon_a^{QD}$ (%) | $\varepsilon_c^{QD}$ (%) | $\varepsilon_a^B$ (%) | $\varepsilon_c^B$ (%) | $\lambda_{RT}$ (nm) |
|---|---|---|---|---|---|
| Without facets | −1.70 | 0.61 | 0.78 | −0.80 | 340 |
| With facets | −1.68 | 0.56 | 0.78 | −0.80 | 349 |

**Figure Captions**

**Figure 1.** Energy loss as a function of depth by an electron beam penetrating in (a) GaN, (b) $Al_{0.67}Ga_{0.33}N$, and (c) AlN for various accelerating voltages, $V_A$. The curves were obtained by performing Monte Carlo simulations using the CASINO software.

**Figure 2.** (a) SEM image of as-grown sample S1A. (b-c) HAADF-STEM views of a nanowire, where bright contrast corresponds to GaN and dark contrast corresponds to AlN: (b) general view of the heterostructure, and (c) zoomed image in the centre of the heterostructure. Dashed yellow lines outline the {1−102} facets in one of the GaN dots. (d) Integrated intensity of the HAADF-STEM image analysed along the turquoise and red rectangles outlined in (c). (e) HAADF-STEM image of the first periods of the heterostructure. Dark/bright contrast corresponds to Al-rich/Ga-rich areas.

**Figure 3.** (a) HAADF-STEM image of sample P1G showing the 88 periods of GaN/AlN. Dark/bright contrast corresponds to Al-rich/Ga-rich areas. (b) Bright-field TEM off-axis off-axis (10° tilt from the [11−20] zone axis) image of P1G showing five periods in the centre of the superlattice, tilted viewed along <1−100>. Bright/dark contrast corresponds to Al-rich/Ga-rich areas. The arrows mark atomic steps at the heterointerfaces. (c) HAADF-STEM image of sample P1N showing the 88 periods of GaN/AlN. Dark/bright contrast corresponds to Al-rich/Ga-rich areas. (d) Bright-field TEM off-axis off-axis (10° tilt from the [11−20] zone axis) image of P1N showing five periods in the centre of the superlattice, viewed along <1−100>. Bright/dark contrast corresponds to Al-rich/Ga-rich areas. The arrows mark atomic steps at the heterointerfaces.

**Figure 4.** (a) XRD θ−2θ scans of samples S1A, P1G and P1N, recorded around the (0002) reflection of GaN. The scans are vertically shifted for clarity. Labels indicate the (111)

reflection of the Si substrate, the (0002) reflection of GaN, and the (0002) reflection of the multi-quantum-well (MQW), with several satellites. (b) Low-temperature PL spectra of samples S1A, P1G and P1N. Spectra are normalised to their maximum value and vertically shifted for clarity (c) Variation of the integrated PL intensity as a function of temperature for samples S1A, P1G and P1N. Measurements are taken under continuous-wave, low-injection (power density = $0.00013$ kW/cm$^2$) excitation. (d) Variation of the IQE at room temperature as a function of the excitation power density measured with a pulsed Nd-YAG laser.

**Figure 5.** (a,b) Schematic description of the simulated structures (a) without taking into account the top facets of the quantum dots, and (b) including the facets (outlined with a red ellipse). (c, d) Left: Band diagram along the [000−1] nanowire growth axis showing three quantum dots in the centre of the superlattice, including a projection of the squared wavefunctions of the first electron ($e_1$) and hole ($h_1$) levels in the central dot. Right: In-plane view of the squared wavefunctions of $e_1$ and $h_1$. Dark contrast indicates higher probability to find the electron/hole. The red hexagons represent the GaN core and the AlN shell. Results in (c) and (d) correspond to the simulated structures described in (a) and (b), respectively. (e,f) Component of the electric field along the [11−20] axis represented along the nanowire diameter. Green/orange curves are obtained at the location of the maximum of the electron/hole wavefunction. The arrows highlight the areas with a change of sign of the electric field if we compare figures (e) and (f). Results in (e) and (f) correspond to the simulated structures described in (a) and (b), respectively.

**Figure 6.** CL spectral line-scan on the cross-section of samples (a) P1G and (b) S1A measured at 7 K. The intensity is colour-coded on a logarithmic scale. The sketches on the left side of the images show the schematics of the samples. White dashed lines highlight the evolution of the peak emission wavelength of the superlattices along the growth axis.

**Figure 7.** CL measurements of samples (a) P1G and (b) S1B as a function of the acceleration voltage. Spectra are normalized at the peak value and vertically shifted for clarity. The emissions from the quantum dot/well superlattice (SL) and from GaN are identified. (c) Variation of the CL intensity as a function of time, measured with $V_A$ = 5 kV and an injection current of 25 pA concentrated in a 20 nm spot. (d) Variation of the CL intensity from sample S1A as a function of the current injected in the sample. Measurements were performed using an electron gun operated in direct current mode, under normal incidence, with a beam spot diameter of 4±1 mm, and keeping $V_A$ = 5 kV.

**Figure 8.** Room-temperature emission from the samples under study normalized to their maxima. The thickness of the AlGaN wells is indicated in the figure. The spectra are vertically shifted for clarity. The spectra from samples with 0.65 nm wells were measured by CL. The rest of the spectra were obtained from PL measurements. Inset: Variation of the IQE as a function of the peak emission wavelength. Different colours represent different content of Al in the wells. The dashed line is a guide for the eye.

**Figure 9.** (a) SEM image of as-grown sample S8. (b-d) HAADF-STEM views of a nanowire, where dark/bright contrast corresponds to Al-rich/Ga-rich areas: (b) general view of the heterostructure, (c) zoomed image of the topmost periods of the heterostructure, and (d) the lowermost periods of the heterostructure.

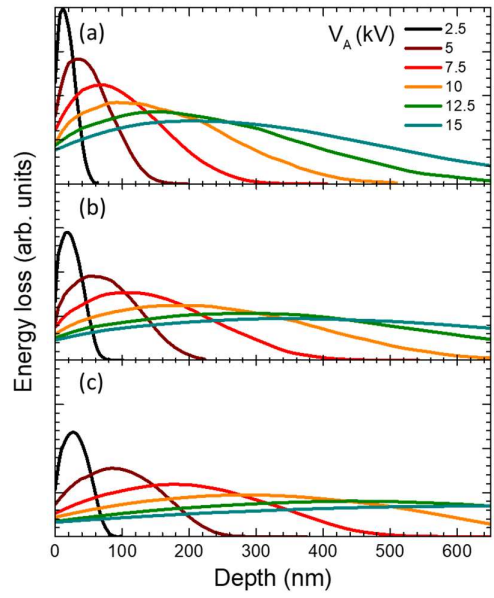

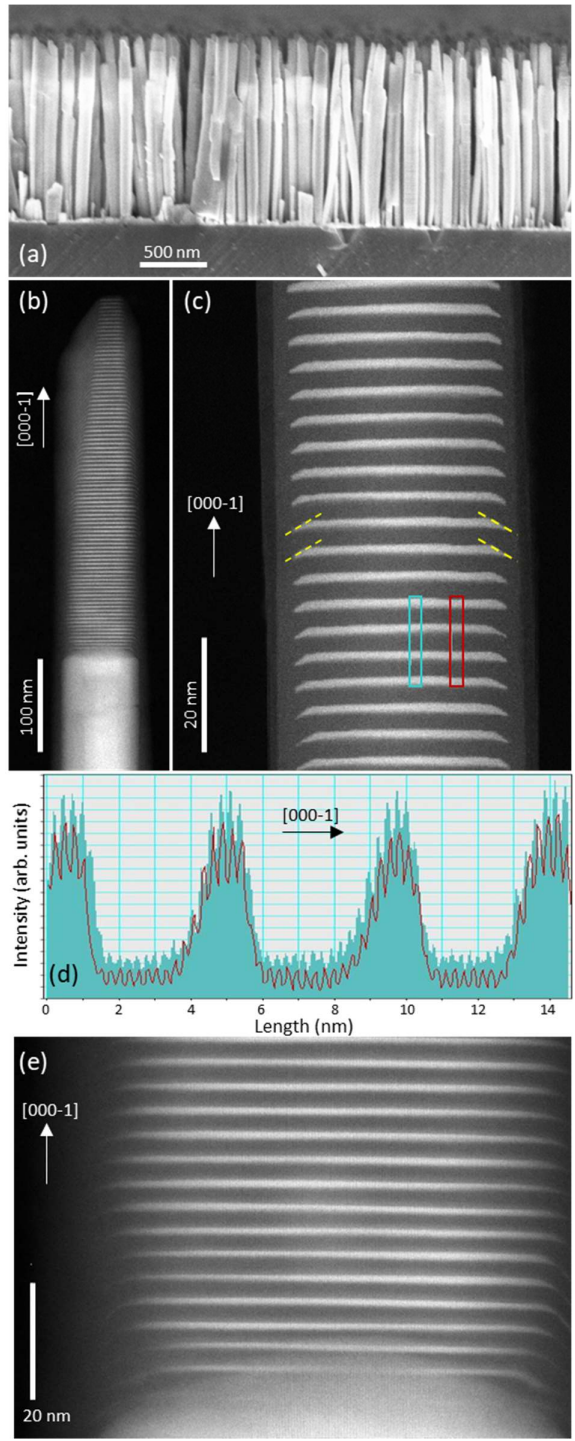

**Figure 3**

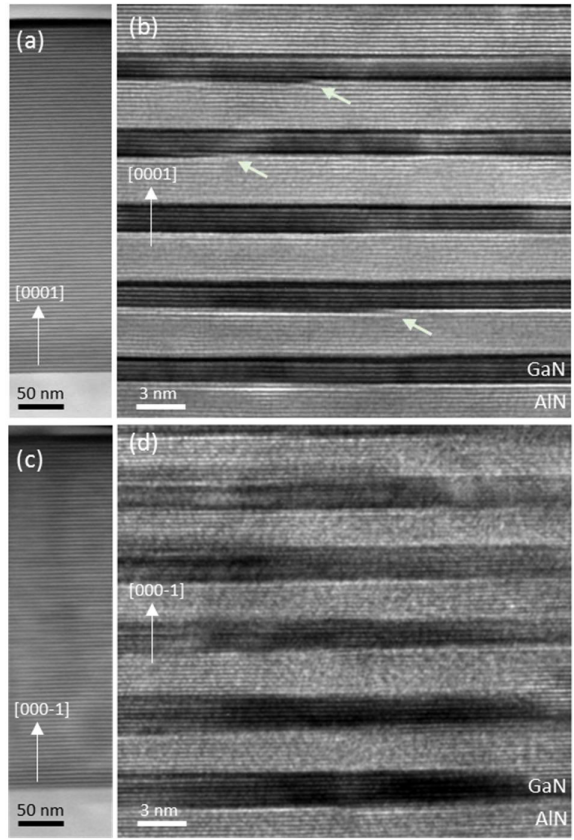

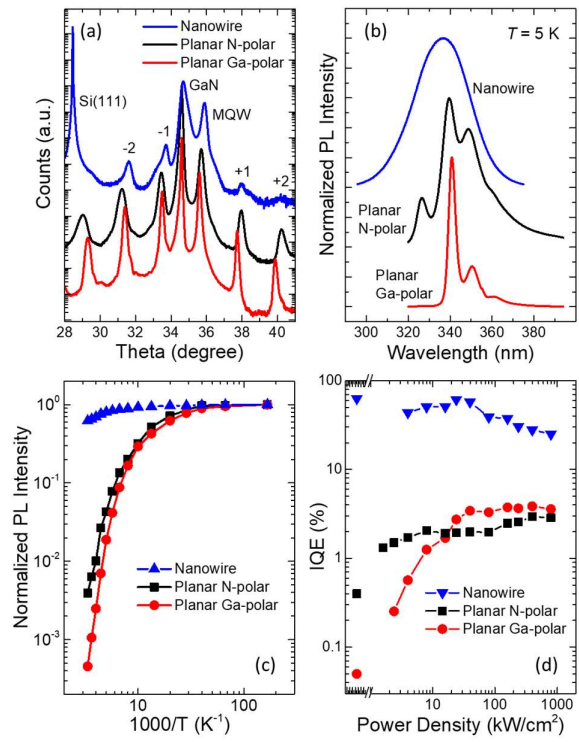

**Figure 5**

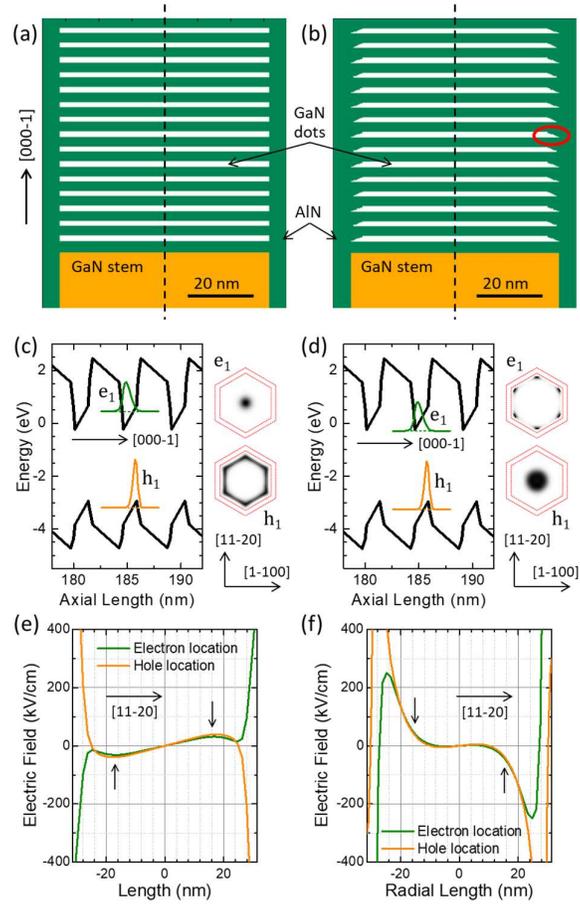

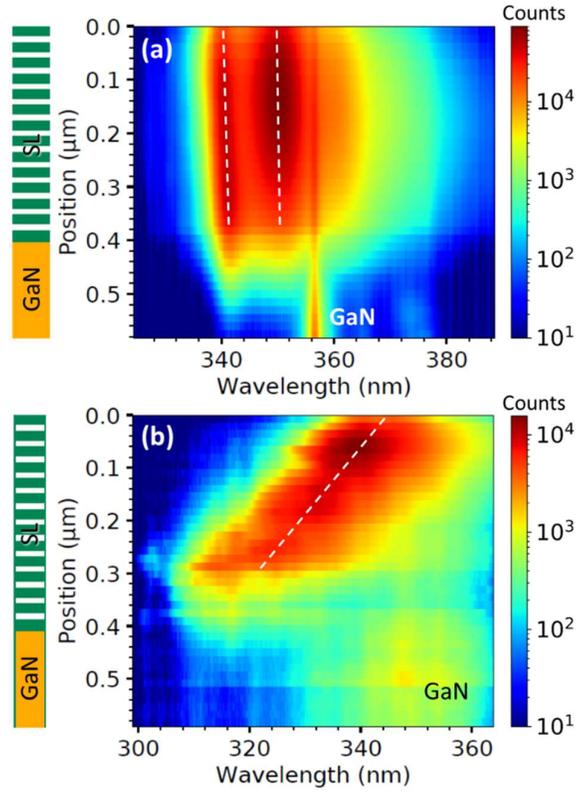

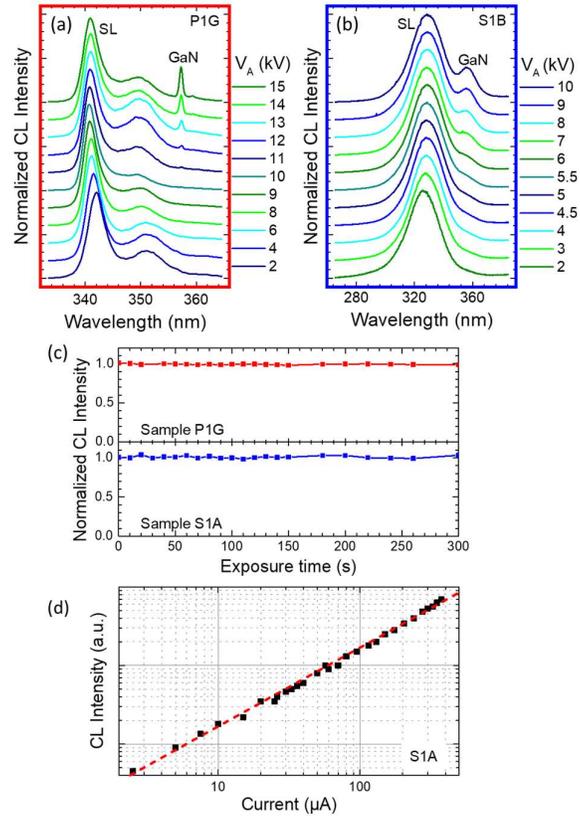

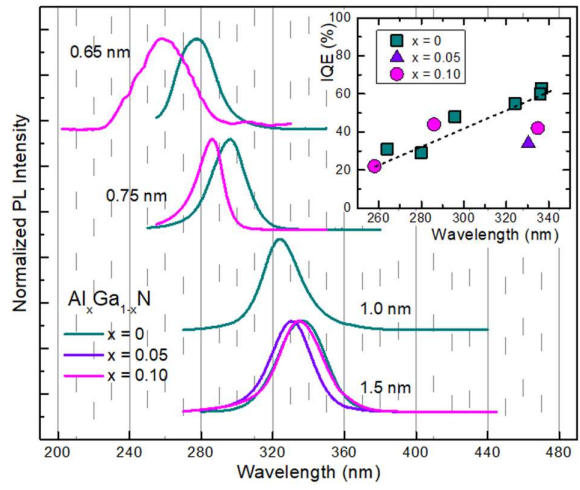

**Figure 9**

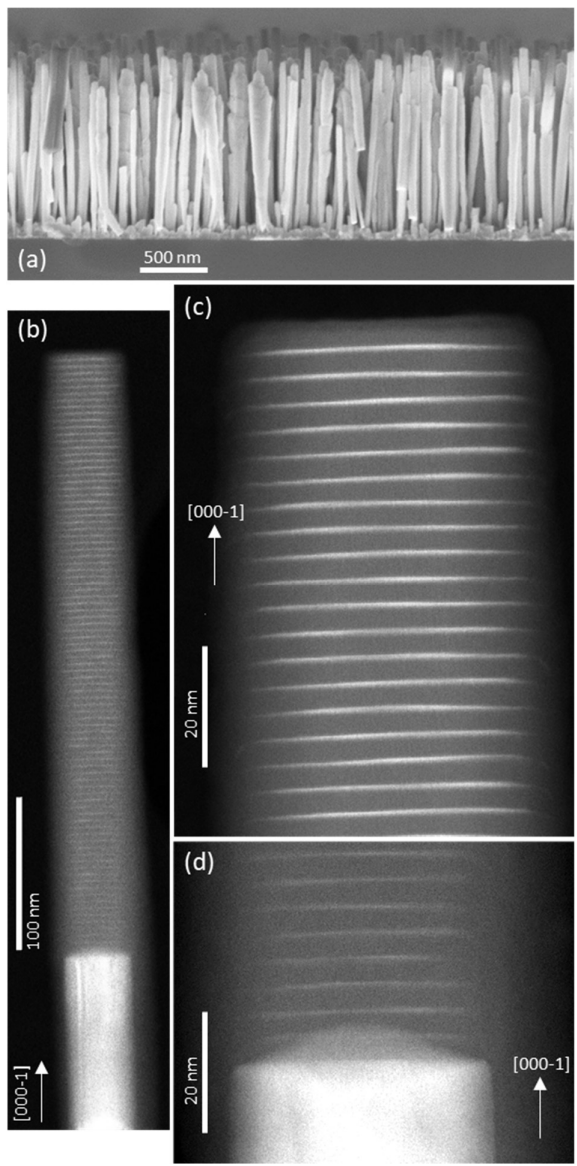